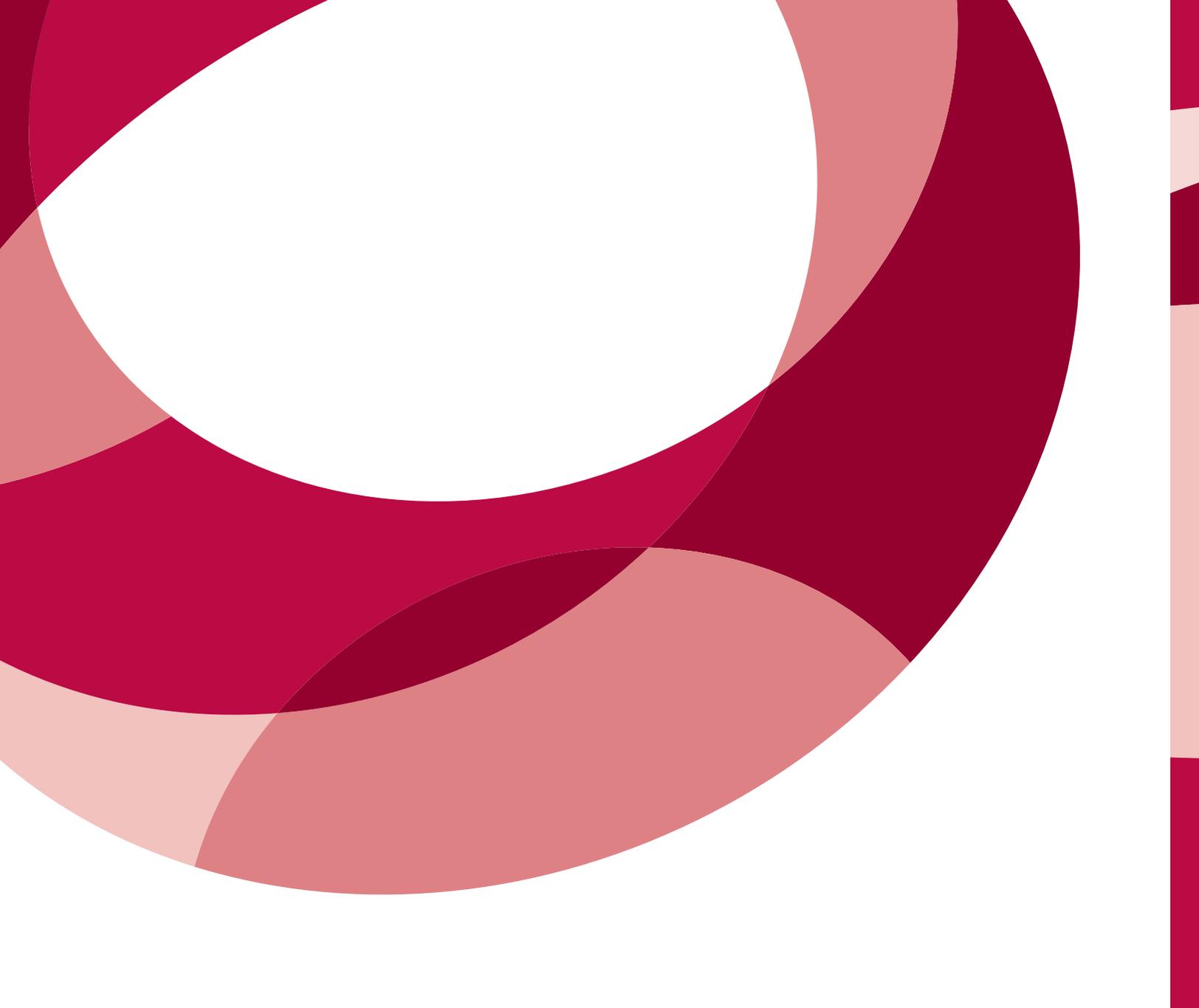

# Thermodynamic Computing

**A REPORT BASED ON A CCC WORKSHOP HELD ON JANUARY 3-5, 2019**

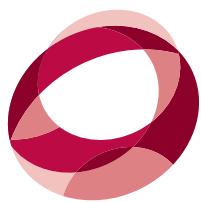

CCC
Computing Community Consortium
Catalyst

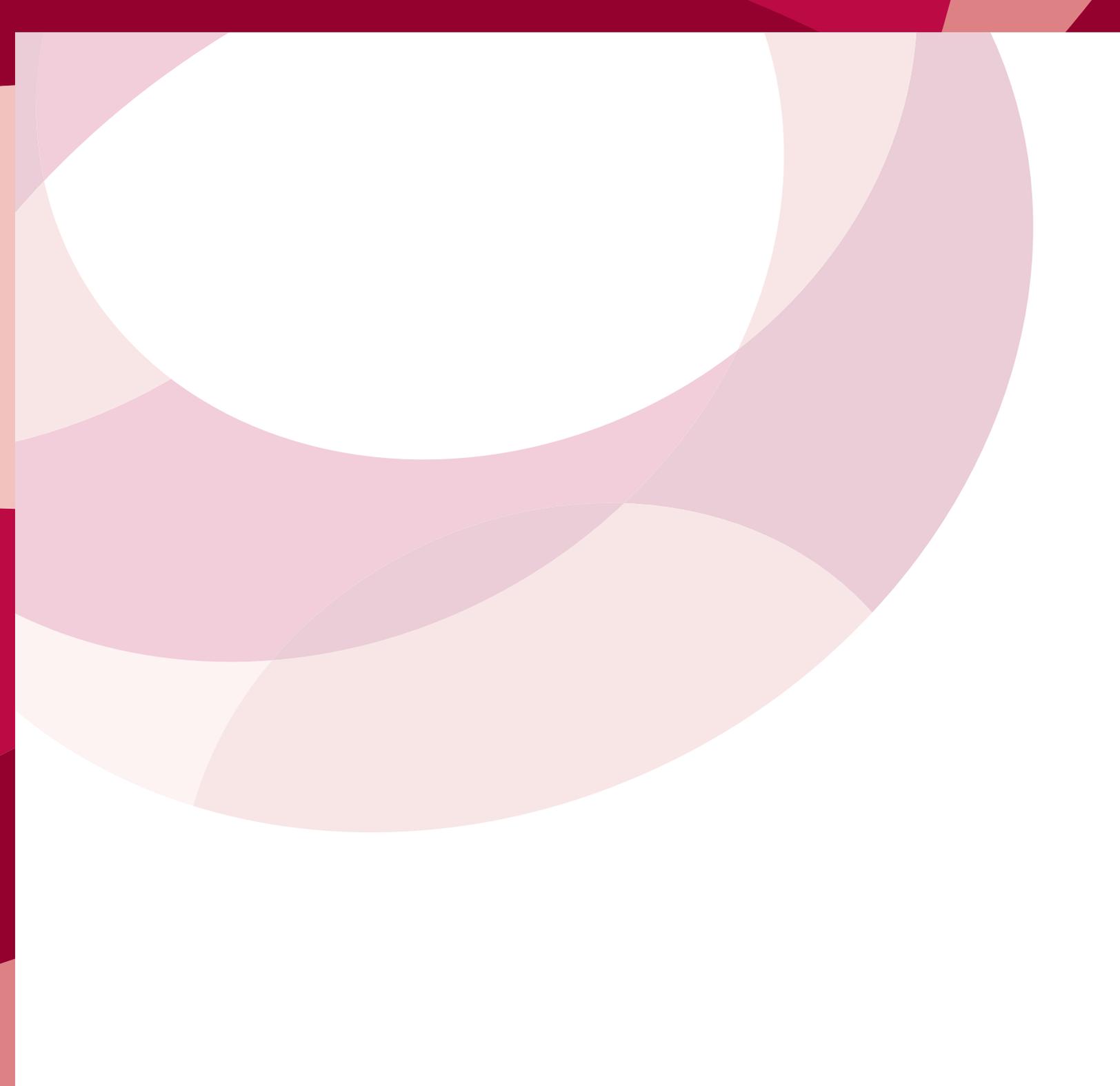

This material is based upon work supported by the National Science Foundation under Grant No. 1734706. Any opinions, findings, and conclusions or recommendations expressed in this material are those of the authors and do not necessarily reflect the views of the National Science Foundation.

# Thermodynamic Computing

**A REPORT BASED ON A CCC WORKSHOP HELD ON JANUARY 3-5, 2019**

**Workshop Organizers:**

Tom Conte, Erik DeBenedictis, Natesh Ganesh, Todd Hylton, Susanne Still, John Paul Strachan, and R. Stanley Williams

**Workshop Participants:**

Alexander Alemi, Lee Altenberg, Gavin Crooks, James Crutchfield, Lidia del Rio, Josh Deutsch, Michael DeWeese, Khari Douglas, Massimiliano Esposito, Michael Frank, Robert Fry, Peter Harsha, Mark Hill, Christopher Kello, Jeff Krichmar, Suhas Kumar, Shih-Chii Liu, Seth Lloyd, Matteo Marsili, Ilya Nemenman, Alex Nugent, Norman Packard, Dana Randall, Peter Sadowski, Narayana Santhanam, Robert Shaw, Adam Stieg, Elan Stopnitzky, Christof Teuscher, Chris Watkins, David Wolpert, Joshua Yang, and Yan Yufik

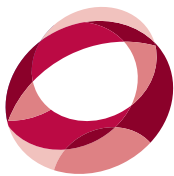

CCC
Computing Community Consortium
Catalyst

# Table of Contents

# 1. Overview and Motivation

## 1.1 Introduction

The hardware and software foundations laid in the first half of the 20th Century enabled the computing technologies that have transformed the world, but these foundations are now under siege. The current computing paradigm, which is the foundation of much of the current standards of living that we now enjoy, faces fundamental limitations that are evident from several perspectives. In terms of hardware, devices have become so small that we are struggling to eliminate the effects of thermodynamic fluctuations, which are unavoidable at the nanometer scale. In terms of software, our ability to imagine and program effective computational abstractions and implementations are clearly challenged in complex domains like economic systems, ecological systems, medicine, social systems, warfare, and autonomous vehicles. Machine learning techniques, such as deep neural networks, present a partial solution to this software challenge, but we hypothesize that these methods, which are still limited by the current paradigm, are a modest subset of what nature does to solve problems. Somewhat paradoxically, while avoiding stochasticity in hardware, we are generating it in software for various machine learning techniques at substantial computational and energetic cost. In terms of systems, currently five percent of the power generated in the US is used to run computing systems — this astonishing figure is neither ecologically sustainable nor economically scalable. Economically, the cost of building next-generation semiconductor fabrication plants has soared past $10 billion and thereby eliminated all but a few companies as the sources of future chips. All of these difficulties — device scaling, software complexity, adaptability, energy consumption, and fabrication economics — indicate that the current computing paradigm has matured and that continued improvements along this path will be limited. If technological progress is to continue and corresponding social and economic benefits are to continue to accrue, computing must become much more capable, energy efficient, and affordable.

We propose that progress in computing can continue under a united, physically grounded, computational paradigm centered on thermodynamics. In some ways this proposition is obvious — if we want to make computers function more efficiently then we should care about energy and its ability to efficiently create state changes — i.e. we should care about thermodynamics. Less clear, but even more compelling, is the long-lingering proposition that thermodynamics drives the self-organization and evolution of natural systems and, similarly, that thermodynamics might drive the self-organization and evolution of future computing systems, making them more capable, more robust, and less costly to build and program. As inspiration and motivation, we note that living systems evolve energy-efficient, universal, self-healing, and complex computational capabilities that dramatically transcend our current technologies. Animals, plants, bacteria, and proteins solve problems by spontaneously finding energy-efficient configurations that enable them to thrive in complex, resource-constrained environments. For example, proteins fold naturally into a low-energy state in response to their environment.[1] In fact, all matter evolves toward low energy configurations in accord with the Laws of Thermodynamics (see Box 1 on page 2). For near equilibrium systems these ideas are well known and have been used extensively in the analysis of computational efficiency and in machine learning techniques. Herein we propose a research agenda to extend these thermodynamic foundations into complex, non-equilibrium, self-organizing systems and apply them holistically to future computing systems that will harness nature's innate computational capacity.

We call this type of computing "Thermodynamic Computing" or TC. Figure 4 (in Section 1.5) illustrates how one can view TC as intermediate between successful Classical Computing and emerging Quantum Computing (this idea is explained in more detail in Section 1.5). At least initially, TC will enable new computing opportunities more than replace Conventional Computing at what Conventional Computing does well (enough) following the disruption path articulated by Christensen (Christensen 2013). These new opportunities will likely be ones that require orders of magnitude more energy efficiency and the ability to self-organize across scales as an intrinsic part of their operation. These may include self-organizing neuromorphic systems and the simulation of complex physical or biological domains, but the history of technology shows that compelling new applications often emerge after the technology is available.

---

[1] Even this relatively simple system is still too compute intensive to model effectively on our most powerful supercomputers — what costs nature a few eV (electronvolts) may cost few TJ (terajoules) on a computer.

## 1.2 A Brief History of Thermodynamics and Computation

The first and last introduction to thermodynamics that most engineering students receive often comes as an introductory first-year course with content that is portrayed as fixed over many decades. It appears to be an exercise in accounting that was defined by dead scientists from centuries ago for the design of steam engines, which focuses on narrow 'equilibrium' conditions using relatively simple mathematics but largely non-intuitive concepts.

The student learns about the four "laws of thermodynamics" and a particularly mysterious quantity known as "entropy" that not only has something to do with "heat" and "free energy" but is also related to the "number of microstates, $W$, available in a system," given by the equation carved into Ludwig Boltzmann's tombstone, $S = k \log W$. Generally, the topics are considered to be unrelated to computation or computing systems, which are never in an equilibrium state. Furthermore, it often seems that everything that can be known about the field was discovered well over a century ago and there is nothing new to learn. However, these perceptions are far from the truth.

What is not well appreciated is that the conceptual foundations of thermodynamics and computing were developed over the same period of time and often by the same researchers, with advances in one domain often inspiring advances in the other. For example, Claude Shannon borrowed the term 'entropy' from thermodynamics to describe uncertainty in information systems because the underlying concepts were so similar. Also, Landauer's seminal work on heat dissipation associated with irreversible logical operations derives from the thermodynamic paradox known as "Maxwell's demon" — an agent supposedly capable of circumventing the second law through the use of information about the system state, which was first described by James Maxwell. Table 1 (page 3) is a partial timeline of the history of thermodynamics and computation. The theme of this report, and the workshop that motivated it, is that this trend continues, with new challenges and opportunities to unite thermodynamics and computing in a new, transformational paradigm. For example, connecting computation to recent theoretical advances in non-equilibrium thermodynamics (Jarzynski 1997; G. E. Crooks 1999) known as "fluctuation theorems" may create substantial new opportunities.

---

### What is Equilibrium Thermodynamics?

Equilibrium thermodynamics is the study of transfers of matter and/or energy in systems as they pass from one state of thermodynamic equilibrium to another, where "thermodynamic equilibrium" indicates a state with no unbalanced potentials, or driving forces, between macroscopically distinct parts of the system. An important goal of equilibrium thermodynamics is to determine how the equilibrium state of a given system changes as its surroundings change.

### Laws of Thermodynamics

**Zeroth Law** – If two systems are each in thermal equilibrium with a third, then they are in thermal equilibrium with each other (or 'there is a game');

**First Law** – Energy cannot be created nor destroyed, but only change forms (or 'you can't win');

**Second Law** – The entropy of an isolated system not in equilibrium will tend to increase over time, approaching a maximum value at equilibrium (or 'you can't break even'); and,

**Third Law** – As temperature approaches absolute zero, the entropy of a system approaches a constant minimum (or 'you can't quit the game').

Box 1: What is Equilibrium Thermodynamics?

| Year | Name(s) of the Creator(s) | Summary of Concept |
|---|---|---|
| 1824 | Sadi Carnot | Description of a reversible heat engine model driven by a temperature difference in 2 thermal reservoirs: "Carnot Cycle." |
| 1837 | Charles Babbage | Specification of the first general-purpose computing system, a mechanical system known as the "Analytical Engine." |
| 1865 | Rudolf Clausius | Definition of entropy and the first and second laws of thermodynamics. |
| 1867 | James Maxwell | Articulation of a thought experiment in which the second law of thermodynamics appeared to be violated: "Maxwell's demon." |
| 1871 | Ludwig Boltzmann | Statistical interpretation of entropy and the second law of thermodynamics. |
| 1902 | Josiah Gibbs | Authoritative description of theories of thermodynamics, statistical mechanics and associated free energies and ensembles. |
| 1905 | Albert Einstein | Theory of stochastic fluctuations displacing particles in a fluid: "Brownian Motion." |
| 1926 | John B. Johnson, Harry Nyquist | Description of thermal fluctuation noise in electronic systems: "Johnson Noise." |
| 1931 | Lars Onsager | Description of reciprocal relations among thermodynamic forces and fluxes in near equilibrium systems: "Onsager Relations." |
| 1932 | John von Neumann | Developments of ergodic theory, quantum statistics, quantum entropy. |
| 1936 | Alan Turing | Description of a minimalistic model of general computation: "Turing Machine." |
| 1938 | Claude Shannon | Description of digital circuit design for Boolean operations. |
| 1944 | Claude Shannon | Articulation of communications theory; foundations of information theory; connection of informational and physical concepts of entropy. |
| 1945 | John von Neumann | Description of computing system architecture separating data and programs: the "Von Neumann Architecture." |
| 1945 | J. Presper Eckert, John Mauchly | Construction of the first electronic computer used initially for the study of thermonuclear weapons: "ENIAC." |
| 1946 | Stanislaw Ulam, Nicholas Metropolis, John von Neumann | First developments of Monte Carlo techniques and thermodynamically inspired algorithms like simulated annealing. |
| 1951 | Alan Turing | Explanation of the development of shapes and patterns in nature: "Chemical Morphogenesis." |
| 1951 | Herbert Callen, Theodore Welton | Articulation of fluctuation-dissipation theorem for systems near equilibrium. |
| 1955 | Ilya Prigogine | Description of dissipation driven self-organization in open thermodynamic systems: "Dissipative Structures." |
| 1957 | E.T. Jaynes | Articulation of the maximum entropy / statistical inference interpretation of thermodynamics: "MaxEnt." |
| 1961 | Rolf Landauer | Explanation of the thermodynamic limits on erasure of information (or any irreversible operations): "Landauer Limit." |
| 1982 | John Hopfield | Description of a model neural network based on the Ising Model: "Hopfield Network." |
| 1987 | Geoffrey Hinton, Terry Sejnowski | Development of a thermodynamically inspired machine learning model based on the Ising Model: "Boltzmann Machine." |
| 1997 | Christopher Jarzynski | Development of an equality relation for free energy changes in non-equilibrium systems: "Jarzynski Equality." |
| 1999 | Gavin Crooks | Development of an equality that relates the relative probability of a space-time trajectory to its time-reversal of the trajectory, and entropy production. This implies Jarzynski equality. |
| 2012 | Alex Krizhevsky, Ilya Sutskever, Geoffrey Hinton | Demonstration of deep machine learning technique in modern computer vision task: "AlexNet." |

*Table I: A timeline of prominent advances in the fields of thermodynamics and computing over the last nearly 200 years. Persistent themes across the centuries include ideas of machines, statistics, fluctuations, and organization.*

## 1.3 Current State of Computing

For a period of time in the 1980s and 1990s, computing was in a stasis where one instruction set (the Intel x86 instruction set) and one operating system (the Microsoft Windows OS) dominated. At the same time, Dennard Scaling — the observation by Robert H. Dennard that as transistors reduced in size due to lithography advances and cleaner processes, the circuits designed with them would increase in speed — dominated the semiconductor roadmap (Dennard et al. 1974). Earlier in 1965, Gordon Moore, co-founder of Intel corporation, published an observation that the economics of the semiconductor industry appeared to follow the relationship of doubling the number of transistors per dollar every 12 months (in 1975, he revised this to every 24 months; hence the commonly used factor of every 18 months, or the average of the two rates) (Moore 1965; Moore and Others 1975). But as with any phenomenological observation of a complex industry, this "law" began to erode. First, as circuits miniaturized in the mid 1990's, wire delays began to dominate transistor speed. In reaction to this, the industry adopted techniques to execute multiple instructions in parallel (e.g. the Intel P6), thereby "hiding" the wire latency and continuing the expectation that computer performance would double every 18 months. For many in the popular press, the combination of Moore's Law, Dennard Scaling, and microarchitectural advances also came to be called, incorrectly, "Moore's Law."

In 2005, Dennard Scaling reached its end, and the power densities of transistor circuits became so large that cooling the microprocessor began to dominate the cost and performance of a system. This new limit was deemed to be the "power wall" that effectively killed the advancement of microprocessor performance. However, the original (and still valid) Moore's Law continued to bring more transistors per dollar every 18 to 24 months. Microprocessor vendors had a choice to make about how to continue: should they invest in sophisticated cooling technologies such as liquid nitrogen or should they use the new abundance of transistors in another way? The choice of the industry was the latter. The extra transistors were used to put more than one processing element (redubbed "core" by Intel marketing) on a single die. However, to use these new microprocessors to speed up applications programmers would need to re-code them using parallel algorithms.

Gene Amdahl, in a now-famous observation from 1965, said that exploiting parallelism is limited by the non-parallelizable portion of a computation (Amdahl 1967). As such, the multi-core approach to microprocessor acceleration in the late 2000s through 2010s could not keep up with the historic doubling of microprocessor performance every 18 to 24 months. As the news of this slowing of compute power began to propagate throughout all corners of the computing community, the Institute of Electrical and Electronics Engineers (IEEE) launched its Rebooting Computing Initiative tasked with rethinking computation from devices to algorithms. One observation of this initiative is shown below in Figure 1 (page 5), which shows the effect of four different approaches to rebooting computing on the various levels in the computing stack. As the approaches moves from left to right in the figure, the degree of disruption is higher and so is the impact on energy efficiency. Thermodynamic computing is decidedly a "level 4" approach to computation and will likely disrupt every level of the current computing architecture.

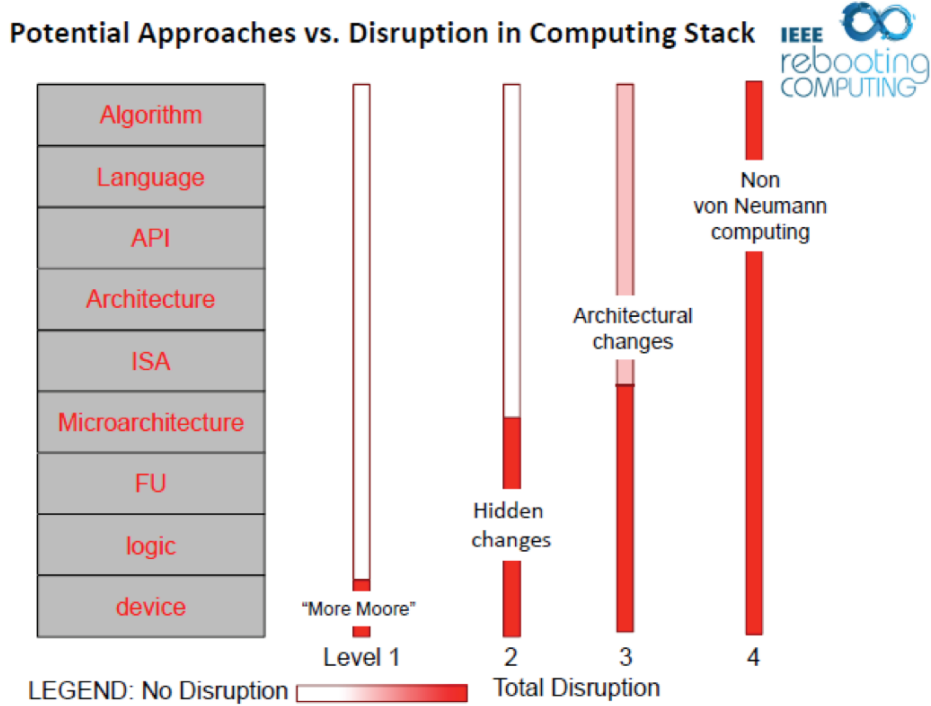

*Figure 1: The different approaches to future computing and their relative required disruption in the computing stack (IEEE Rebooting Computing Initiative, used with permission).*

## 1.4 Recasting Computing in Thermodynamic Terms

To begin thinking about computing in terms of thermodynamics, it is instructive to consider what the stack model shown in Figure 1 assumes with respect to the physical systems that implement it. At each level in the stack small-scale details are coarse-grained[2] to present higher-level features to superior levels.[3] In addition, within each level "components" or "modules" are engineered such that their small-scale dynamics are isolated from one another. For example, electronic circuit components interact through coarse-grained electrical signals and the small-scale dynamics in different circuit components are disconnected. This allows us to think about circuit elements as "transistors" or "resistors." These circuits can then be "coarse-grained" to engineer higher-level logic gates in which the smaller scale dynamics of the circuits within the gates are independent. This allows us to think about gates as components needed to engineer higher-level computing elements like Arithmetic Logic Units. These same ideas apply to software systems in which software levels comprised of various modules present abstract interfaces to higher software levels and protect the internal details of their modules. This allows us to think of software in terms of "drivers," "libraries," "operating systems," "applications," etc. This strategy of coarse-graining, layering, and isolating components within a computing system is driven by our need to understand, engineer, and program them. It is the enabling foundation of the current conceptual paradigm in computing in which we think of computers as state machines that implement mathematical operations, but it comes with large costs because it ignores the underlying thermodynamics.

---

[2] *Coarse-graining* is the process of abstracting out many small-scale details in order to create simple representations at a larger scale. For example, when we buy gasoline we care about the coarse-grained concept of "gallons" and not about the details of what all the molecules in the gas are doing. See Section 2.2.

[3] This idea is also known as *scale separation* in physics. The central concept is that most of the details at smaller scales of organization are irrelevant at higher scales. Unlike engineered systems, living systems cannot be well understood with ideas of coarse graining and scale separation – they are so called "complex" or "multiscale" systems. See Section 2.2.

Recasting computing in thermodynamic terms begins with the realization that all processes in the physical world are driven by the dissipation of free energy. Computing is also a physical process, but the current paradigm views computation as a kind of mathematical or state transition process. The reality, however, is that computation is a carefully engineered, deterministic sequence of state transitions that dissipate free energy. In the current paradigm, thermodynamics is viewed only as an engineering constraint motivating energy efficient hardware designs and effective heat removal.[4] The limitations of today's paradigm are evident when we apply even basic thermodynamic considerations. As an example, we note that there are many different ways a computer can be organized in order to implement a particular function but that each of these ways has different thermodynamic properties. For example, the thermodynamic cost of an AND gate — i.e. the drop in entropy over its states when it runs, which must ultimately be dissipated as heat when the gate is reset — will depend on the distribution of the states on its inputs, since that determines its initial entropy. Since this distribution will depend on the location of the gate in the larger system, the global layout of the system will affect the thermodynamic costs of the gates from which it is comprised — even if the overall system implements the same logical function. Clearly then, there are thermodynamic implications that are not captured in the current computing paradigm.

## 1.5 A Vision for Thermodynamic Computing

We envision a Thermodynamic Computer (TC) as an engineered, multi-scale, complex system that, when exposed to external potentials (inputs), spontaneously transitions between states in the short term while refining its organization in the long term, both as part of an inherent, adaptive process driven by thermodynamics. Like quantum computers, TCs are distinguished by their ability to employ the underlying physics of computing substrate to accomplish a task. For example, TCs may employ naturally occurring device-level fluctuations to explore a state space and to stabilize on low-energy representations, which may then be employed in an engineered task. In this section, we outline a vision for thermodynamic computing and in Section 2 (page 13) we describe the scientific challenges and research directions needed to realize this vision.

As illustrated in Figure 2 (page 7), a thermodynamic computing system (TCS) is a combination of a conventional computing system and novel TC hardware. Like many heterogeneous systems, the conventional computer is "host" through which users can access the TC. As with all conventional computers, humans control every aspect of the host computer and are its "interface" to the real world. The TC, on the other hand, is directly connected to real-world potentials, which drive the evolution of its internal organization. The host computer can be thought of as providing constraints on the TC that select the external potentials of interest and configure/program some portions of the thermodynamic hardware. In a TCS *computing constrains thermodynamics*, inverting the *thermodynamics constrains computing* perspective of the current paradigm. Figure 3 (page 8) is a conceptual schematic of a TC and its environment. The TC is imagined as a number of evolvable "cores" or "elements" that spontaneously transition to low energy states depending on their inputs. The cores are embedded in a network of evolvable connections, and a subset of the cores is connected to the environment. Like all computing systems, the environment is a collection of electrical and information potentials.[5] Environmental potentials drive currents into the TC, which, if they cannot be effectively transmitted back to the environment (by connecting positive and negative potentials, for example), must be dissipated within the TC hardware. This dissipation creates fluctuations of the system state that can be stabilized if they are more effective at transmitting charge/reducing dissipation. This adaptation, if successful at multiple spatial and temporal scales, can result in a system with complex, multiscale dynamics.[6] Among existing computing systems, TC is most similar to neuromorphic computing, except that it replaces rule-driven adaptation and neuro-biological emulation with thermo-physical evolution.

---

[4] We refer to the analysis of thermodynamic constraints in conventional computing as the "thermodynamics of computation," which we distinguish from "thermodynamic computing."

[5] Unlike conventional computers, the use of information potentials requires their translation to and from electrical potentials – likely a kind of digital-to-analog and analog-to-digital conversion. Currently this translation would appear to be more of a conceptual challenge than an engineering challenge as the connection between information and thermodynamics has not been completely established.

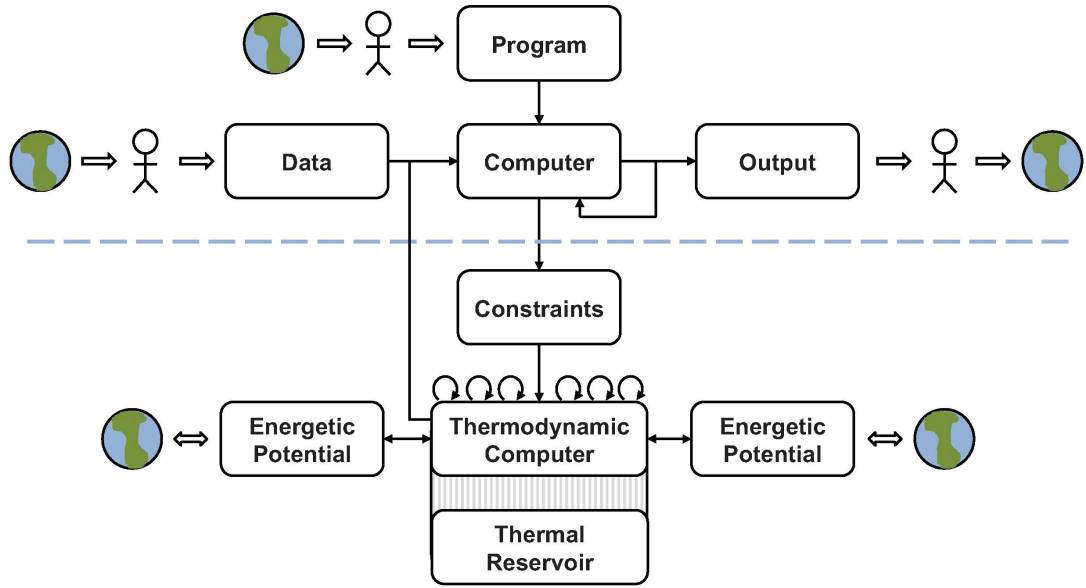

*Figure 2: Conceptual schematic for a Thermodynamic Computing System. The top half of the figure represents a conventional computing system that "hosts" the TC. The host computer is entirely prescribed by humans, who are its interface to the real world. The TC, illustrated in the lower half of the figure, has independent interfaces to raw potential in its environment and complex, multiscale, recurrent adaptive internal evolution that communicate and connect environmental potentials. Humans can direct the evolution of TCs by programming constraints that influence the connections to the environment and the evolution of the system. The TC can also provide feedback to the conventional computing system; for example, it may evolve a representation of its environment that can be used as input. The thermal reservoir plays an active role in the evolution of the TC by providing fluctuations.*

As an example of using a TCS, a user might program constraints that describe an optimization objective over the external potentials and that capture the TC's natural tendency to maximize entropy production in the environment while minimizing entropy production internally (an idea that we address further in Section 2.1). In other words, the user sets up the TC so that its thermodynamics solve an optimization objective of interest.

To put this vision in a larger context, Figure 4 (page 8) divides computing into domains according to their relationship to fluctuation scales. Spatial and temporal fluctuation scales are estimated in terms of thermal energy (kT) and corresponding electronic quantum coherence times and lengths. We divide the computing paradigm into three qualitatively different domains that we label as "Classical," "Thermodynamic," and "Quantum."

**Classical Domain:** In the classical domain, fluctuations are small compared to the smallest devices in a computing system (e.g. transistors, gates, memory elements), thereby separating the scales of "computation" and "fluctuation" and enabling abstractions like device state and the mechanization of state transformation that underpin the current computing paradigm. The need to average over many physical degrees of freedom in order construct fluctuation-free state variables is one reason that classical computing systems cannot approach the thermodynamic limits of efficiency. Equilibrium thermodynamics and Newtonian physics are also in this classical domain and we are able to define macroscopic thermodynamic state descriptions with concepts of temperature, pressure, entropy, free energy,[7] and macroscopic mechanical properties like mass, hardness, and flexibility because fluctuations are irrelevant at large scales. The classical computing paradigm fails when fluctuations approach feature size, which is a central challenge in computing today.

---

[6] Many of these ideas are illustrated in a recently developed model called Thermodynamic Neural Network (Hylton 2019).

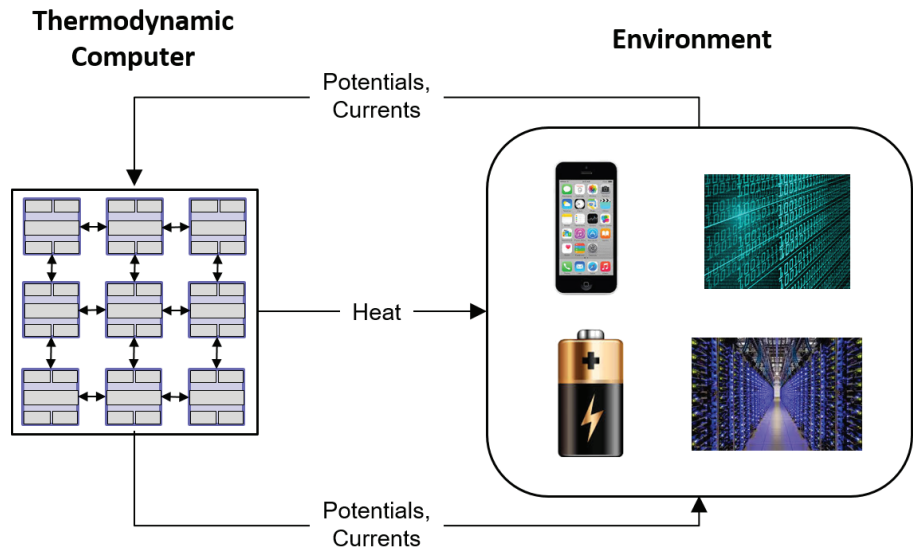

*Figure 3: Conceptual schematic for Thermodynamic Computing and its environment. A TC can be thought of as a generic fabric of thermodynamically evolvable elements or cores embedded in a network of reconfigurable connections. External potentials drive the flow of electrical currents through the network. Energy dissipation in the TC creates fluctuations in the system state. Fluctuations that decrease dissipation are spontaneously stabilized. Correspondingly, the TC evolves to move current through the network with minimal loss as it equilibrates with its environment. The environment comprises various information, electrical potentials, and a thermal reservoir.*

**Quantum Domain:** In the quantum domain, fluctuations in space and time are large compared to the computing system. While the classical domain avoids fluctuations by "averaging them away," the quantum domain avoids them by "freezing them out" at very low temperatures (milli-Kelvin for some systems). In the quantum domain we see many interesting and non-intuitive effects of quantum mechanics and the potential for a qualitatively different computing capability as compared to the classical domain.

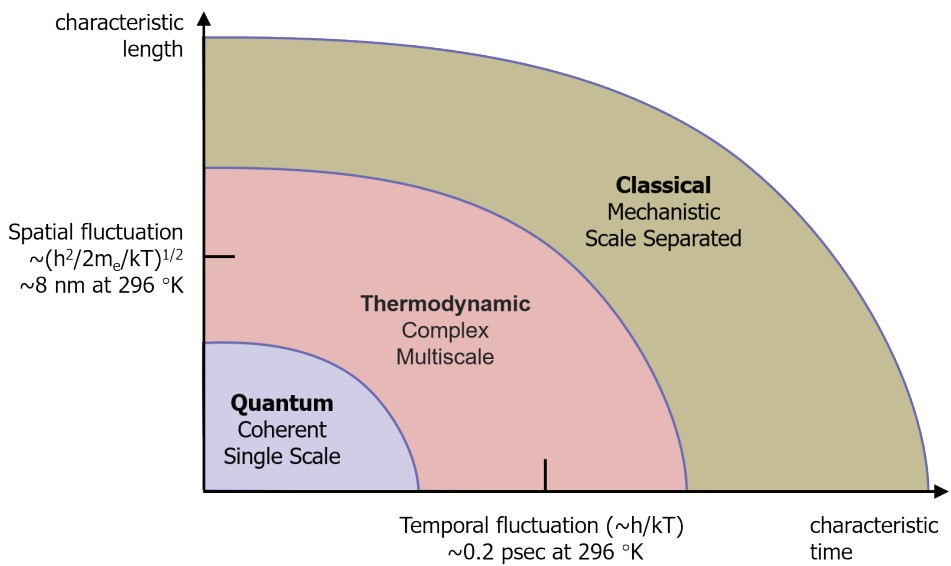

*Figure 4: The three major domains of computing*

---

[7] The term "thermodynamic limit" is used to describe this scale-separated regime of large equilibrium thermodynamic systems. We prefer the term "classical limit," as it applies to large, scale-separated systems more generally.

**Thermodynamic Domain:** In the thermodynamic domain, fluctuations in space and time are comparable to the scale of the computing system and/or the devices that comprise the computing system. This is the domain of non-equilibrium, mesoscale thermodynamics and statistical physics. It is also the domain of cellular operations, neuronal plasticity, genetic evolution, etc. — i.e. it is the domain of self-organization and the evolution of life. The thermodynamic domain is inherently multiscale and fluctuations are unavoidable. Presumably, this is the domain that we need to understand if our goal is to build technologies that operate near the thermodynamic limits of efficiency and spontaneously self-organize, but it is also the domain that we carefully avoid in our current classical and quantum computing efforts.

Figure 5 is another illustration of these same ideas emphasizing the relevant ideas when considering systems near room temperature. Atoms and small molecules are in the quantum domain, the components that are used to build computers are in the classical domain, and the components of life (and future computing systems) are in the thermodynamic domain.

While there is much that is still to be understood, engineering a thermodynamic computer will require that its state space, fluctuations, and adaptability suit its intended domain of input potentials, as well as an interface that allows a programmer to define the problem to solve by configuring in advance some portions of the TC. With these ideas in mind, we foresee the following "TC Roadmap" for the development of Thermodynamic Computing systems.

### 1.5.1 TC Roadmap

Below we describe three broad development stages (with substages) that we believe will be required in the transition to thermodynamic computing.

**Stage 1A:** Model System Development: Today's computing systems can be used to model future stages on the TC Roadmap. Noteworthy existing models include the Boltzmann Machine (Hinton, Sejnowski, and Others 1986), and some of today's machine learning research is clearly relevant to TCs of the future and can provide insight into how they should operate. The Thermodynamic Neural Network model (Hylton 2019) is a recent contribution with this objective clearly in mind.

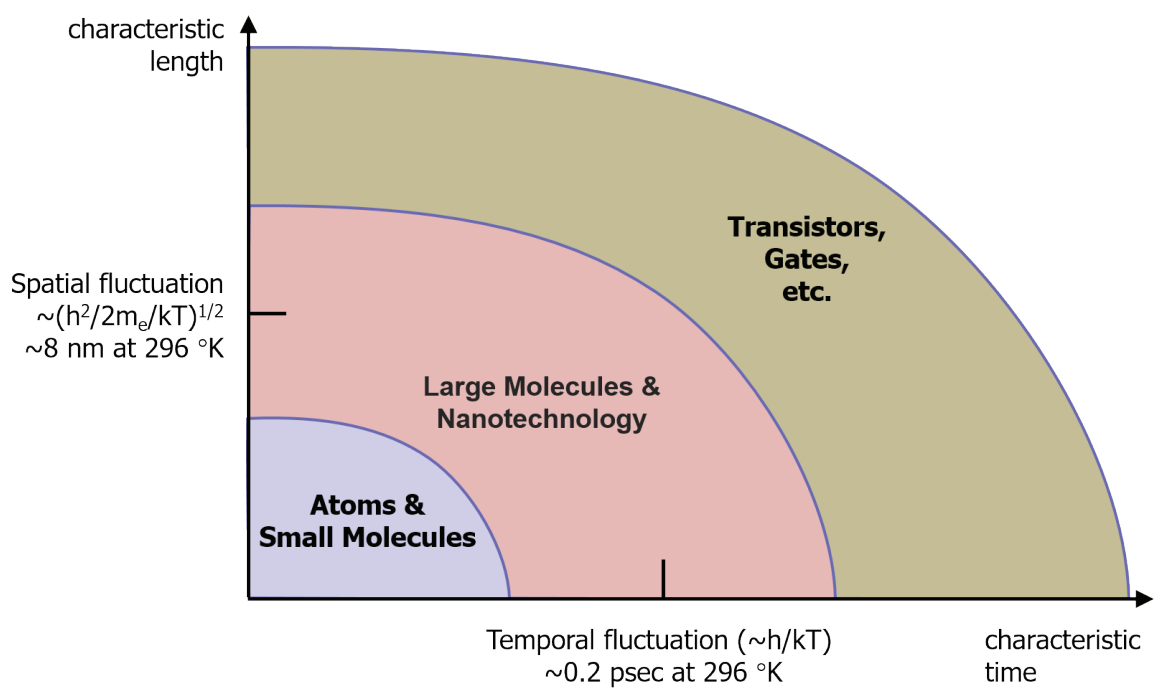

*Figure 5: Comparison of fluctuation scales and characteristic sizes of physical, biological, and computing systems.*

**Stage 1B:** Thermodynamic Optimization of Classical Computing: As argued in Section 1.4, thermodynamic efficiency is more complex than what is captured in current ideas of computational complexity. The details of the state transformations bear upon the efficiency of computation in ways that are not yet captured in current hardware, compilers, or runtimes. Thermodynamic optimization seeks to improve the performance of existing computing systems through the consideration of fundamental thermodynamic costs regarding the sequence of state transformations at every level of the computing stack. Understanding these costs may not only improve the efficiency of today's computing paradigm but should also create insights into future stages of the roadmap.

**Stage 2:** Thermodynamically Augmented Classical Computation: The classical computing paradigm is joined to collections of relatively simple thermodynamic elements representing evolvable state variables within the classical computing system. As thermodynamic elements evolve, they may transition between fluctuation dominated operation and classical operation through the shedding or accumulation of microscopic components (e.g. charge, filaments) from which state variables are constructed and presented to the classical system. Fluctuations are physical but they are confined conceptually and operationally as isolated components that are linked together through a classical computing system/network. Example components might include thermodynamic "bits," "neurons," "synapses," "gates," and "noise generators." This domain is likely accessible today as combinations of conventional computing systems and novel electronic/ionic components such as memristors. This domain will likely have strong conceptual ties to chemistry and materials science. After achieving some success with these augmented systems it could be possible to develop a full-fledged TCS,

**Stage 3A:** Complex Thermodynamic Networks: A classical computing system provides an interface to and scaffolding for mesoscale assemblies of interacting, self-organizing components exhibiting complex dynamics and multiscale, continuously evolving structure. State variables emerge and decay spontaneously as collections of semi-stable structures in response to external potentials. This domain may become accessible as we gain experience with thermodynamically augmented classical systems and as we continue to reduce the scale of fabricated components deep into the nanometer regime. This domain will likely have strong conceptual ties to domains such as neuroscience, cell biology, molecular biology, and mesoscale, non-equilibrium, and statistical physics.

**Stage 3B:** Quantum Thermodynamic Networks: Coherent domains of quantum computing elements (e.g. Q-bits) are coupled to other quantum domains through quantum-thermodynamic fluctuations, enabling the evolution of hybrid quantum-thermodynamic-classical systems. Like quantum computers, these systems will likely operate at very low temperatures and connect to the external world through a classical computing system. This domain may become accessible as we gain experience with complex thermodynamic networks and quantum computing systems. This domain will likely have strong conceptual ties to quantum and statistical physics.

## 1.6. Summary of Scientific Challenges and Research Directions

Here we summarize the primary scientific challenges and research directions identified during the workshop. Many of the ideas presented in this section are discussed in greater detail in Section 2 (page 13) of this report.

### 1.6.1 Scientific Challenges

From a physical perspective, and at the most basic level, our understanding of non-equilibrium thermodynamics and self-organization in complex systems remains incomplete. As a salient example, modern deep learning methods, which we see as precursors to TCs, have demonstrated success on a wide range of tasks, but we still lack a deep understanding of why and how they work. The overarching scientific challenge is to connect and refine our understanding of dissipation, fluctuation, equilibration, and adaptation in non-equilibrium, open thermodynamic systems. While we understand each of these in limited contexts and observe their obvious unification in living systems, our shortcomings are evident when considering the challenge of building TCs. We further understand that our current computing technologies are far from the limits of thermodynamic efficiency, but we don't yet know how these limits might be approached if this fundamental challenge were successfully addressed.

From a computational perspective, and at the largest level, the scientific challenge is to develop a new computing paradigm on which to ground the development of thermodynamic computing technology. A host of challenges seeking to modify and expand the current paradigm will emerge including the development of (1) thermodynamic computational complexity classifications, (2) thermodynamic computing components for Stage 2 systems that can be assembled into architectures that generate diverse behaviors, (3) models, measures, and tools to describe thermodynamic adaptation and efficiency, and (4) an overarching framework describing a thermodynamic computing system comprised of an evolving system of components, inputs and outputs, energy dissipation pathways, a heat bath, and an optimization goal or computational task defined by a human.

These physically and computationally inspired perspectives are complemented by a number of scientific challenges that couple them together. Examples of these challenges include (1) specifying algorithms and architectures employing noisy, energy-constrained, evolving components and their interactions, (2) understanding the role of naturally generated randomness and its effect on probabilistic algorithm implementation and self-organization, (3) understanding the multiscale interactions among the architectural levels and components of a TC, (4) understanding how to program, train, and evolve a TC, and (5) characterizing and quantifying the performance of a TC including the role of the various components, architectures, tasks, etc.

### 1.6.2 Research Directions

With these scientific challenges in mind, we foresee the following research directions that should be pursued to address them: (1) core theoretical research, (2) model systems, (3) building blocks, and (4) TC system architectures.

Core theoretical research is needed to expand our current understanding of non-equilibrium thermodynamics of complex open systems. These efforts may well build upon existing work in fluctuation-dissipation theorems (Jarzynski 1997; G. E. Crooks 1999) and the relationship between thermodynamics and prediction (Still et al. 2012). In particular, a formalism that describes the phenomena of adaptation and self-organization is needed. Such formalisms might be applied to existing optimization and machine learning methods, and they may also be tested against small physical systems similar to those used to test non-equilibrium fluctuation-dissipation theorems (Collin et al. 2005). Although such research should be transformative in many fields, in the context of thermodynamic computing it will likely (1) inform the development of the thermodynamic computing paradigm, (2) elucidate the thermodynamic underpinnings of existing optimization and machine learning methods, (3) derive performance bounds of TCs, including trade-offs among speed, accuracy, energy dissipation, memory usage, physical size, memory stability, and effects of noise, and (4) inform the development of thermodynamic computing architectures and components.

In addition to this core theoretical work, additional theoretical research is needed to develop model systems that elaborate and demonstrate these ideas. These efforts may build upon existing model frameworks like Hopfield nets (Hopfield 1982), Boltzmann Machines (Hinton, Sejnowski, and Others 1986), thermodynamic neural computation models (Fry, 2005), and Thermodynamic Neural Networks (Hylton 2019) or they may also abstract from biological systems. The goal of these efforts will be to (1) serve as testbeds to evaluate core theoretical concepts, (2) construct proof-of-concept implementations for representative problems to be addressed by TCs, and (3) provide guidance for the development of TC systems and components.

Enabling materials, devices, and components research is necessary to provide the building blocks for TC systems. The overarching motivation is to increase functionality in a given volume of total material through complex, thermodynamic interactions rather than increasing the areal density of separated components. Potential starting points for this work include ongoing research efforts in Josephson junctions, memristive systems, magnetic nanopillars, single electron systems, etc. The goals of this work are (1) to develop new classes of devices that reconfigure thermodynamically and are sufficient to enable complex TC operation, (2) to create mathematical models of device behavior, and (3) to expand current design, synthesis, and simulation tools models to include exposure of the thermodynamic variables, not just voltages and currents.

Finally, research into the design, construction, and evaluation of TC architectures is needed to synthesize conceptual and component research into useful systems. The likely starting points for much of this work are the tools and methods supporting the design of computers today. Research directions in system development include developing (1) languages to describe TC systems including problem-description languages, programming languages, hardware specification languages, system specification languages, and self-organization languages, (2) specification and simulation framework for prototyping TCs, and (3) benchmark tasks and evaluation metrics.

## 1.7. Summary of Social and Technological Impacts

If the scientific challenges and research directions above are successful, the impact on science will be both profound and broad. It is rare in human history to be able to articulate the need for a new fundamental understanding while also having the means to begin its exploration. The opportunity described in this report is the result of a convergence of difficulties in the current computing paradigm driving the need for the creation of a new paradigm that will not only support the future of computing, but also the future of our understanding of the natural world.

### 1.7.1 Scientific Impacts

From a computational perspective, the development of a thermodynamically centered computation paradigm would:

- Enable computation near fundamental limits of efficiency;
- Enable self-organization within the computing system without dependence on a human specified program;
- Enable probabilistic, stochastic, reversible, and machine-learning computing concepts to be united in a common paradigm;
- Expand the set of physical devices and components that can be used for computation, by enabling the use of inherently probabilistic and/or unreliable components; and
- Enable computing systems to connect to and become part of the real world because they are physical systems from top to bottom.

From a broader perspective, the development of the conceptual foundations for thermodynamic computing would:

- Enable understanding of the organization and computational power of living systems, potentially including the spontaneous emergence of "intelligence";
- Enable humans to better mimic the extraordinary capabilities of living systems in human engineered systems; and
- Enable a substantive convergence of concepts from diverse fields in engineering, physical sciences, biological sciences, and social sciences.

From the perspective of grand challenge problems, the development of large-scale thermodynamic computing systems would:

- Enable access to much larger, more capable, and more adaptable computing resources at a much lower cost; and
- Address a broad class of computational modeling problems that are currently intractable in domains like biology, medicine, ecology, climate, and social and economic systems.

### 1.7.2 Social and Technological Impacts

From the perspective of the ongoing improvement in computing systems, the development of thermodynamic computing systems would:

- Enable dramatic decreases in energy consumption per unit of computational work performed;
- Enable dramatic increases in software development efficiency by offloading much of the detail of the system organization to the thermodynamic computer;
- Enable large increases in battery life for portable and edge connected computing systems; and
- Enable a very large increase in the capabilities of small, low-cost, computing systems, such as perceptual capabilities that rival those of animal sensory systems.

From the perspective of impact on society in general and on the US in particular, the development of thermodynamic computing technologies would:

- Sustain US leadership in emerging computational paradigms;
- Develop a uniquely capable, cross disciplinary workforce;

- Lower the environmental impact of computing systems;
- Increase access to computing technologies and services;
- Improve outcomes in most human enterprise, including medicine, business, agriculture, defense, security, leisure, etc.; and
- Create new and broad classes of transformative business and social opportunities with dramatic impact.

## 2. Overarching Themes
### 2.1 Theory

**BACKGROUND AND SUMMARY OF THE CURRENT STATE**

As has been the case throughout history, theoretical physics has a central role in defining the next generation computing paradigm and technology; in this case thermodynamic computers. Today computing is driven by a decidedly non-physical paradigm — symbol manipulation and rule-driven state transformation — but the implementation of any computing systems is ultimately still physical. The current situation is not unlike that faced by the many pioneers summarized in the timeline of Section 1.2 (page 2); similarly, we see an opportunity and need to develop the foundations to realize it.

Elementary "demons" that control thermodynamic systems using "intelligence" were introduced long ago (Maxwell, 1888) (Leff & Rex, 2002) by way of highlighting thermodynamics' iron-fisted Second Law of increasing disorder, with a most-prescient analysis completed by Szilard via his single-molecule engine (Szilard, 1929). The net result, as emphasized by him, was that these constructions are at best consistent with the Second Law. That is, these devices produce no net thermodynamic gain. Curiously, and despite Szilard's insights, much controversy about their functioning persisted in the following decades. A fuller understanding started to emerge in the 1960s when the central role of information gained recognition, largely through Landauer (1961), Penrose (1970), and Bennett (1982). A well-known result of these works is Landauer's conclusion that each bit of information erased at temperature $T$ results in at least $k_B T \ln(2)$ of energy dissipation. At the simplest level, this follows directly from the second law of thermodynamics and the statistical-mechanical understanding of entropy as unknown information. Whenever some previously known (i.e., correlated) information is lost (e.g. to the environment) and is subsequently thermalized, with its prior correlations having become inaccessible, total entropy has increased and free energy has decreased.

**LIMITATIONS OF THE CURRENT APPROACH**

Our conventional *irreversible* approach to digital computation loses information all the time — in particular, every time a circuit node (which could be the output node of a logic gate, a memory cell, or an interconnect line) is destructively overwritten with a new logic value the information contained in the old value is lost. Furthermore, conventional implementations of digital logic lose far more information than this, because typically a large amount of (redundant) physical information is used to encode each bit of digital information. Even at the projected end of the semiconductor roadmap in ~2033, each typical digital bit contains on the order of 60,000 $kT$ of electrical energy which is dissipated as heat when that bit is erased, while the smallest circuit components (transistor gates) store only about 100 $kT$ of electrical energy, which is close to the limit for reliable operation set by thermal noise, so something fundamental will have to be done differently in order to continue improving the energy efficiency of computation.

From a higher-level perspective, the fundamental problem that the current computing paradigm cannot address is that the computers cannot organize themselves except in carefully engineered contexts in which an optimization objective and method can be defined. We can see this limitation reflected in the many people who are needed to build and maintain software systems. Machine learning addresses these challenges in limited contexts, but we have no paradigm for generic self-organization in computing systems. These challenges are discussed in Section 2.2 (page 18).

**INSIGHTS**

Thermodynamics was originally developed to understand the energetic efficiency of heat engines: large, macroscopic machines for converting temperature gradients into mechanical work. The second law of thermodynamics places fundamental limits on the efficiency of such engines. The work performed on a system must be no less than the net change in its free energy: W ≥ ΔF. The dissipation-free limit can only be approached for machines operating quasi-statically where the transformations are slow and the machines remain in thermodynamic equilibrium. In the past two decades, however, we have witnessed a fundamental shift in our understanding of the laws governing the thermodynamics of microscopic systems operating far from equilibrium. These ideas have set the stage for the development of a new thermodynamic paradigm for computation.

### 1. Reversible Computing

Reversible computing (that is, computation without losing information) is perhaps the earliest conceptualization of thermodynamic computing. Considered by Landauer (1961) in his original paper, reversible computing was proven to be Turing universal by Bennett (1982), and Bennett and Landauer collaborated over many years to develop the fundamental thermodynamics of reversible computing in more depth. In reversible computing, information within the machine is transformed *reversibly* and some fraction of the signal energy associated with that information can be recovered and reused for multiple computational steps without being dissipated to heat. Reversible computing is that aspect of thermodynamic computing that makes the smallest departure from computing as practiced today. It is still digital (not analog), still deterministic (not stochastic), still explicitly engineered (not self-organizing), and still focused on traditional algorithmic, pre-programmed computation (not on learning).

### 2. Fluctuation Theorems

Jarzynski (1997) showed that the second law inequality, $W \geq \Delta F$, can be derived from a simple equality, $\langle exp[-\beta W] \rangle = exp[-\beta \Delta F]$ where $\beta$ denotes inverse temperature and the angular brackets represent an average over many repetitions of the process. The Crooks' fluctuation theorem (Crooks 1998) establishes the relationship between the relative likelihoods of different dynamical paths or trajectories that the microstates of a non-equilibrium system could traverse and the entropy production associated with those trajectories, due to this it is an important milestone in the field. Using time-reversal symmetry and conservation of energy, Crooks derived the following relationship

$$\frac{\pi(\gamma)}{\pi(\gamma^*)} = exp\left[\frac{\Delta Q(\gamma)}{k_B T}\right]$$

where the left hand side is the ratio of the relative likelihoods of a certain trajectory ɣ of microstates (a sequence of microstates over time) to it's time reversed trajectory ɣ*, and $\Delta Q(\gamma)$ is the heat dissipated into the thermal reservoir (at temperature T) as the system traverses the trajectory shown in the figure below. The relationship above indicates that a certain forward trajectory is more likely than the time reversed one by an exponential factor of the heat $\Delta Q(\gamma)$. The relationship is extremely powerful as it holds even in the presence of external fields driving the system.

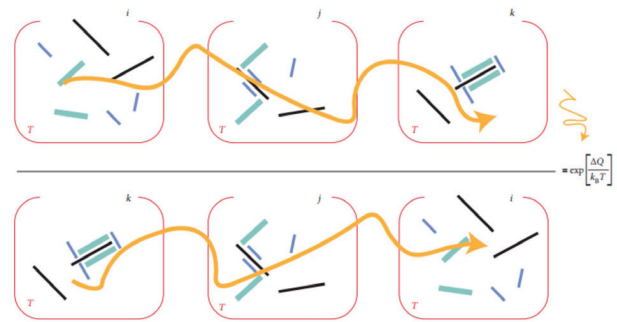

*Figure 6: The Crooks Fluctuation theorem provides a quantitative relationship between the likelihoods of the forward and reverse trajectory of microstates when driven by an external field with the heat dissipated Q into the thermal bath as the system traverses the trajectory. This figure originally appeared in Jeremy England's* Dissipative adaptation in driven self-assembly *(England 2015).*

While Crooks' Fluctuation theorem dealt with microtrajectores in systems, England (England 2015) (Kachman, Owen, and England 2017) recently developed a generalization of this relationship to understand the likelihood ratios associated with transition between macrostates (shown in figure 7 below) by integrating over all microstates under a macrostate and over all relevant trajectories. These were used to study the thermodynamic conditions for the emergence of intelligence – adaptive learning *(reliable high dissipation)* and predictive inference *(reliable low dissipation)* in self-organized systems.

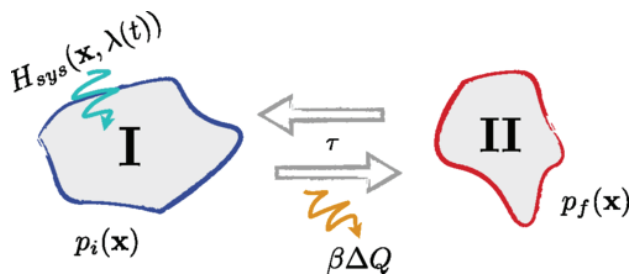

*Figure 7: The macrostate fluctuation theorem quantifies the relationship between the likelihood of driving a system in macrostate I to macrostate II (in their corresponding microstate distributions) in time τ with the internal entropy change in the system and the heat dissipated into the bath $\Delta Q$. This figure originally appeared in Nikolay Perunov, Robert A. Marsland, and Jeremy L. England's* Statistical Physics of Adaptation *(Perunov et al. 2016).*

Using these fluctuation theorems, Still and collaborators (Still et al. 2012) have shown that systems operating at high thermodynamic efficiency are those that preserve information that enables the prediction of future inputs. England argues that dissipation of absorbed work drives the adaptation of open thermodynamic systems and describes the connection to modern fluctuation theorems. Ganesh (2018) discusses the overarching conceptual and philosophical implications of these theories.

### 3. Thermodynamics of Information and Computation

Recent theoretical ideas, such as the fluctuation theorems just described, have fostered new understanding of the thermodynamics of information and computation and have enabled the development of models and physical principles for what is now called *information thermodynamics* (Sagawa, 2012) (Parrondo, Horowitz, & Sagawa, 2015). Due to recent innovations in experimental technique new examples of elementary devices/demons have now been realized in the lab (Toyabe, Sagawa, Ueda, Muneyuki, & Sano, 2010) (Berut, et al., 2012) (Koski, Maisi, & Pekola, 2014) (Jun, Gavrilov, & Bechhoefer, 2014) (Dechant, Kiesel, & Lutz, 2015) (Koski, Kutvonen, Khaymovich, Ala-Nissila, & Pekola, 2015) (Rossnagel, et al., 2016). A number of closely related thermodynamic costs of classical computing have been identified including the following examples:

◗ The information-processing second law (Boyd, Mandal, & Crutchfield, 2016) (Deffner & Jarzynski, 2013) that extends Landauer's original bound on erasure to dissipation in general computing and properly highlights the central role of information generation measured via the physical substrate's dynamical Kolmogorov-Sinai entropy (Sinai & G., 1959). It specifies the minimum amount of energy that must be supplied to drive a given amount of computation forward. It justifies the conclusion that information is a thermodynamic resource.

◗ Coupled thermodynamic systems, for example a thermodynamic computer in a complex environment, incur transient costs as the system synchronizes to, predicts, and then adapts to fluctuations in its environment (Boyd, Mandal, & Crutchfield 2016b) (Boyd, Mandal, & Crutchfield, 2017) (Still et al. 2012) (Boyd A. B., Mandal, Riechers, & Crutchfield, 2017).

◗ The modularity of a system's organization imposes thermodynamic costs (Boyd, Mandal, & Crutchfield, 2018).

◗ Costs due to driving transitions between information-storage states in far from equilibrium and non-steady state systems (Riechers & Crutchfield, 2017).

◗ Costs of generating randomness (Aghamohammdi & Crutchfield, 2017), which is itself a widely useful computational resource.

Employing these principles, new strategies for optimally controlling non-equilibrium transformations have been introduced (Zulkowski, Sivak, & DeWeese, 2013) (Zulkowski & DeWeese, 2015) (Gingrich, Rotsko, Crooks, & Geissler, 2016) (Sivak & Crooks, 2016) (Patra & Jarzynski, 2017). These are complemented with new diagnostic techniques — the trajectory-class fluctuation theorems (Wimsatt, et al., 2019) — that use mesoscopic observables (such as work) to diagnose successful and failed information processing by microscopic trajectories. These results, summarized in Table 2, describe much of the current state of theoretical understanding in the thermodynamics of information and computing. The consequence of these developments is that we are now poised to implement thermodynamically and computationally functional nanoscale systems.

**IMPLICATIONS FOR THERMODYNAMIC COMPUTING**

A key feature of microscopic dynamics considered in the theories described above is that thermal fluctuations can be large compared to the scale of the system. Consequently, microscopic machines, operating with very low energy budgets, are inevitably stochastic and random. They cannot operate deterministically like clockwork (at least not while also being thermodynamically efficient). For microscopic machines operating at thermal energy scales fluctuations are large and inevitable. In contrast to today's computers, which are built upon stable digital components, thermodynamic computers will be built from noisy components, coupled by linear and nonlinear interactions offering a much higher density of bits, much lower energy dissipation per elementary operation, and operation in the presence of noise.

| Principle | Meaning |
|---|---|
| Information Destruction $$\langle W \rangle \geq k_B T \ln(2)$$ | Logically irreversible operations dissipate energy (Landauer, 1961) |
| Reciprocity $$\langle W_{min}^{t-sym} \rangle = k_B T \langle \Psi \rangle - \langle W \rangle$$ | Logically nonreciprocal operations dissipate energy |
| Information Creation $$\dot{Q} \geq k_B T \ln(2)(h_\mu - h'_\mu/\hat{L})$$ | Creating information dissipates heat (Aghamohammdi & Crutchfield, 2017) |
| Information Process Second Law $$\langle W \rangle \geq k_B T (h'_\mu - h_\mu)$$ | Work to drive (or energy dissipated) during computation (Boyd, Mandal, & Crutchfield, 2016a) |
| Requisite Complexity $$\langle W \rangle \leq k_B T \ln(2) \min\{\Delta H_1, \Delta h_\mu\}$$ | Advantage maximized when controller matches environment (Boyd, Mandal, & Crutchfield, 2016b) |
| Synchronization & Error Correction $$\langle Q^{tran} \rangle_{min} \geq k_B T \ln(2) \, I[X_0 : \bar{Y}'] - E'$$ | Work to correct errors or synchronize to environment (Boyd, Mandal, & Crutchfield, 2017) |
| Modularity $$\langle \Sigma_{t \to t+\tau}^{mod} \rangle_{min} = k_B T \ln(2) \, \Delta I_{t \to t+\tau}$$ | Controller modularity is thermodynamically expensive (Boyd, Mandal, & Crutchfield, 2018) |
| Information Dynamics $$\lambda > 0$$ | Maxwellian demons are chaotic dynamical systems (Boyd & Crutchfield, 2016) |
| Steady-State Transitions $$\Pr(W_{ex}, \Psi)/\Pr(-W_{ex}, -\Psi) = e^\Psi e^{\beta W_{ex}}$$ | Work to drive transitions between information storage states (Riechers & Crutchfield, 2017) |
| Functional Fluctuations $$I(u) = (\beta^{-1} - 1) h_\mu(P_\beta) - \beta^{-1} \log_2(\hat{\lambda}_\beta)$$ | Engine functionality fluctuates in small systems, short times (Crutchfield & Aghamohammdi, 2016) |
| Control tradeoffs $$\dot{\Sigma} = f(1/\tau, L^2)$$ | Counterdiabatic control dissipation design (Campbell & De, 2017) (Boyd, Patra, Jarzynski, & Crutchfield, 2018) |
| Reliability $$\dot{\Sigma} = f(-\ln(\epsilon))$$ | Dissipation costs of high-reliability information processing |
| Trajectory-class fluctuations $$\langle e^{-W/k_B T} \rangle_C = R(C^R)/P(C) \, e^{-\Delta F/k_B T}$$ | Success and failure have thermodynamic signatures (Wimsatt, et al., 2019) |

*Table 2: Nonequilibrium thermodynamics of information processing in classical physical systems. For notation, refer to the cited works.*

The mesoscale in between conventional transistors that operate using millions of electrons and quantum bits that operate at the single electron level is a complex regime in which few-body dynamics result in a rich ecosystem of behaviors. Semiconductor and solid state components with features smaller than a few nanometers exhibit different physical dynamics than their bulked up cousins: few-body interactions give rise to large statistical fluctuations, quantum effects such as tunneling dominate dynamics, and quantum noise introduces novel sources of uncertainty. In this domain the necessity of understanding the fundamental physics of information processing is clear. The thermodynamics of computation enforces minimum rates of dissipation for logically irreversible processes, while non-equilibrium fluctuation theorems relate the speed, energy dissipation, and noisiness of elementary thermodynamic logical processes. While it will take time to realize this potential, current semiconductor devices are already pushing down to the relevant length scales — for example, TSMC's announcement of a 5nm process node.

Physical limits to computation suggest that exascale ($10^{18}$ operations (ops) per second) thermodynamic computers could be attainable in the relatively near future by extending existing technologies with potential for Zetta scale ($10^{21}$ ops per second) and Yotta scale (Avogadro's number of ops per second). An overarching principle for these bounds is that the efficient use of energy requires effective information processing. Recycling of free energy inside the thermodynamic computer will play an essential role. Thermodynamic computers will process and communicate not only information as part of their operation but also energy flow. New theory is needed to understand the deep connections among these concepts for thermodynamic computing. Programming thermodynamic computers raises multiple issues in the fields of optimal and stochastic control, including the control of quantum mechanical systems. In order to design and control thermodynamic computers, we must turn to the same laws of physics that drove us to construct them in the first place.

In light of the considerations just described, we see the following list of scientific challenges (questions to answer) and research directions (developments to undertake).

**Scientific Challenges**

◗ Develop an integrated non-equilibrium theory of fluctuation, dissipation, adaptation, and equilibration that address, for example, long standing problems of stability, noise, quantum effects, reversibility, etc.

◗ Clarify and expand the relationship between information theory, computation, and thermodynamics.

**Research Directions**

◗ Extend non-equilibrium fluctuation theorem development toward the domain of thermodynamic computing;

◗ Develop model systems to support the refinement of thermodynamic computing theory and development;

◗ Characterize existing semiconductor and unconventional computing components near thermodynamic limits where fluctuations are avoidable, then compare results to theoretical predictions; and

◗ Integrate recent theoretical and experimental results on small-scale, fluctuating devices into larger component systems.

## 2.2 Self Organization

**BACKGROUND AND SUMMARY OF THE CURRENT STATE**

Classical statistical physics assumes separation of scales into microscopic and macroscopic domains. Macroscopic thermodynamic quantities are averages over the microscopic domain and most of the microscopic details become irrelevant by this averaging (e.g. fluctuations at microscopic scales). Systems are generally thought of as "ensembles" of instances, each with a different microscopic configuration. Various types of ensembles are used according to certain system-scale constraints that are applied. An important refinement of these ideas was the introduction of renormalization group theory to address multiscale fluctuations in the vicinity of second order phase transitions — a situation in which simple averaging over the small-scale features was insufficient. In all these approaches, the result is a system representation at large-scale using only a small number of parameters — the "sufficient statistics" of the model. Additionally, systems with different microscopic properties can yield identical macroscopic properties because averaging within and across scales eliminates details that are "irrelevant" at large scales, which is often referred to as "universality." For example, the central limit theorem shows that averaging a large number of identical, independently distributed probability distributions results in a Gaussian distribution regardless of the details of the component distributions.

In this classical picture from statistical physics, the second law of thermodynamics is often imagined as a system evolving from a (somehow previously prepared) less probable ensemble of states to a more probable ensemble of states in a simple, closed system. These more probable ensembles are necessarily more "disordered" than the initial distributions from which they evolved leading to the idea that the second law of thermodynamics is exclusively concerned with the creation of disorder. However, this classical, closed system picture is obviously not consistent with our experience of everyday life. Organization spontaneously emerges, evolves, and disappears everywhere because the world is an open thermodynamic system with abundant sources of free energy.

Machine learning concepts are an interesting mix of ideas that derive largely from thermodynamics. Ideas of probability, statistics, stochastics, and energy minimization arguably derive from classical, equilibrium thermodynamic concepts. Ideas of learning, prediction, self-organization, parameter evolution, and error propagation are arguably the domain of non-equilibrium thermodynamic concepts. Not surprisingly, it is this latter set of ideas that has fueled the recent renaissance in machine learning. In thermodynamic parlance, machine learning systems evolve structure via parameter refinement (on a cost/loss/energy function) to exploit (informational) free energy in the data sets on which they are trained.

**LIMITATIONS OF THE CURRENT APPROACH**

A typical computational goal is the development of a model that represents a complex, real-world system of interest. Among other requirements, such models must have at least a similar number of states as the real world system that it seeks to represent — sometimes described as the "law of requisite variety" (Ashby 1991). The complexity and fluidity of real world systems, however, makes it difficult to assess just how this should be done. A common approach is to attempt to reduce the description of the real world system to its "fundamental elements and interactions" and to replicate those elements and interactions with a system of algorithms in a large-scale model. For example, in order to model the macroscopic activity of the brain, this reductionist approach might select neurons as the fundamental elements and attempt to model the brain as a system of neurons. However, both common sense and classical thermodynamics tell us that the vast majority of the details at the scale of the neuron don't matter and that much that is outside the brain does matter regarding its macroscopic activity. By and large, attempts to model real-world systems as very large collections of interacting component-level algorithms are both expensive and ineffective in creating a representation of the macroscopic features of the systems that they seek to portray. Machine learning models circumvent some of this difficulty by refining parameters to better fit the system of interest but the difficulties associated with the reductionist approach still remain.

Complex real world systems can be effectively described neither by simple macroscopic averages (classical thermodynamics) nor by detailed microscopic algorithmic models (reductionist approach) (Bar-Yam 2016). Instead, we hypothesize that they should be described as systems that interact and reconfigure resources across multiple scales in response to environmental inputs. For example, if the environment demands lifting a very heavy object, muscle fibers within and across muscle groups unite in a common task that may be effectively described with a few pieces of information applied uniformly to the muscle fibers. On the other hand, if the environment demands a complex task like playing a musical instrument, fibers within and across muscle groups are recruited for differing, fine scale motions and their description requires relatively more information applied to selected portions of the muscular system. Also, to further complicate this descriptive task, although most small-scale details are irrelevant at the large-scale, in complex systems some small-scale features have large impact at large-scale. For example, a few bits of information from the environment can cause someone to "change his mind," to move over long distances, to quit a job, to become emotional, etc.; similarly, a flawed DNA replication event can lead to a cancer that overwhelms a much larger organism.

Unfortunately, we lack the elegant methods of classical thermodynamics to create low-dimensional macroscopic representations that connect scales and remove irrelevant microscopic features in non-equilibrium open systems. Hence, it is not surprising that the conceptual foundations of computing today do not capture concepts like multi-scale interdependence, complexity, phase transitions, and criticality. But it is also the case that they do not even capture the most basic thermodynamic concepts of energy, entropy, and associated conservation laws. As it seems overwhelmingly likely that these ideas are central to the phenomenon of self-organization it is not surprising that current computing systems have no such inherent capacity. To suppose that there exists an "algorithm of the brain," or "a universal plasticity rule," or "basic cortical microcircuit," or "artificial general intelligence," or "digital twin" is also to suppose that such ideas can be effectively described by humans and captured in software, which seems extremely unlikely given the complexity of the real world systems that we might seek to represent.

**INSIGHT AND HYPOTHESES TO ADDRESS THESE LIMITATIONS**

Life is clearly an example of a self-organizing, multiscale, open thermodynamic system and is frequently used as an "existence proof" of feasibility in the quest for self-organizing technological systems. We hypothesize that living systems are spectacularly energy efficient precisely because their evolution is driven by thermodynamics and that this effect is not limited to their genetics, but applies as well to their development and ongoing adaptation within their environments. Interestingly, we can make the same claim regarding today's computing systems if we also consider humans to be part of that system. Like living systems our joint human-computer systems are also evolving under thermodynamic pressures to increase efficiency, to address competition, and to access new sources of free energy to sustain their ongoing existence and evolution. We use this observation to imagine a future for human-TC systems and to define some of the roles and properties of a TC.

Firstly, we suppose that TCs will always be used as joint human-TC systems, because we will always want these TCs to serve us. We hypothesize that the role of a TC in this joint system is to self-organize across scales to create large-scale representations of value to the humans who will "finish the job" using their unique knowledge of how these representations may be of value to them. This is not unlike the current paradigm in which certain groups (users) rely on other groups (developers) to work out the (small-scale) details to create an application that serves their (large-scale) needs. For example, when we access a mapping application on a smartphone we provide only the large-scale information that is uniquely ours (e.g. where we want to go) while the application works through myriad details to make it happen. The combination of the developers and today's computing systems is effectively a "TC;" so, by examining and abstracting what developers do today, we can understand what TCs should do in the future.

Software system development today is a complex, self-organizing enterprise across many scales and tasks. Tasks vary in size and scope and require a corresponding allocation of development resources. Some tasks are at small-scale (e.g. "bare metal programming") and some are at large-scale (e.g. "data lakes" and "user interfaces"). Subsystems may provide representations to other subsystem as "interfaces" that hide "irrelevant" details within them. Hierarchies may emerge in which performance at the largest scales is largely immune to the details of the small-scale implementations, but because the scales are interconnected some small-scale changes have effect at large-scale (e.g. a "bug"). Also, changes at large-scale (e.g. large changes in usage patterns, denial of service attacks) may break systems at a small scale. In general, the development efforts across scales support, influence, and constrain each other and we often refer to the software system that emerges as a "stack". The evolution of these systems is driven by thermodynamics (typically mediated by money, although sometimes pizza is used instead) and happens in a complex, changing environment with competition, limited resources, and time constraints. The software system is also never "finished" in that its survival depends on constant upgrades, patches, maintenance, performance improvements, feature enhancements, user uptake, money flow, etc.

By analogy with the software development paradigm just described, the task of a TC is to draw effectively upon its internal resources and to create high-level representations of value to its human user. This user will be largely unaware of the many irrelevant microscopic details of how the TC created this representation but can, by his/her interaction with it at large-scale, influence the organization at smaller scale. If the TCs task is to represent a complex real world system, which seems a very likely and desirable task, then it must have sufficient internal resources to create a faithful representation. In general this suggests that a TC be comprised of a very large number of elements that may be grouped and organized across scales to address the complexity inherent in the real world system of interest. We might imagine, for example, an evolving network of interconnected units connected to an external source of potential provided by the real world system and the human user that drives its evolution (some might argue that this what a brain does). Such a system would not be a collection of automata and algorithms specifying their connection; it would be a multiscale, open, non-equilibrium, thermodynamically evolving system. The central challenge for the development of TCs is understanding how to do just this.

In light of the considerations just described, we seek to understand the following scientific challenges and develop the following research directions:

**Scientific Challenges**

◗ Decisively abstracting and describing the phenomenon of self-organization in open, non-equilibrium thermodynamic systems, perhaps as an extension of recent efforts to develop fluctuation-dissipation theorems. While ad hoc and qualitative ideas exist in different domains, none is yet sufficient to engineer a technology.

◗ Understanding the effects of scale and scale-separation on the performance and architecture of a TC. Will a TC be characterized by a separation of long (program-like) and short (data-like) time scale elements? Should multi-scale interactions be engineered as part of an evolving "stack?"

◗ Understanding how a TC should be "trained." For example, should a TC be "shaped" by training component organizations on simple tasks before being jointly trained on more complex tasks? How will prior experience be accessed for new tasks?

◗ Understanding the tradeoffs and roles of humans and TCs in the joint human-TC context. What portions of the system should be accorded to users, developers, and the TC? To what degree and in what ways should a TC be engineered or programmed to sufficiently describe the high-level task of interest and constrain the allowable solution space while leveraging the TCs ability to self-organize?

◗ Creating interfaces between a TC and the real world system that it seeks to represent. How do we translate the real world system features and behaviors into inputs that drive the organization of the TC? For example, can the real world system be sampled as "information" which is then recognized as a thermodynamic potential by the TC?

**Research Directions**

◗ Expand and enhance recent theoretical work in non-equilibrium, multi-scale thermodynamic systems to include understanding of the evolution of organization.

◗ Integrate and unify concepts from information theory, complexity, machine learning, thermodynamics, biology, statistics, dynamical systems, and related domains to improve and expand their descriptive power and to create a common language.

◗ Build model systems, for example in computer simulations or in hardware prototypes, that both illustrate self-organization and promote the development of technological systems. Use these model systems to guide the development of thermodynamic computing design, training methodologies, interfaces, and analysis tools. Natural starting points include Ising models, Hopfield nets, Boltzmann machines, and various biologically inspired models and machine learning systems and techniques.

◗ Create heuristics and/or simple analyses that can effectively constrain the requirements for building a TC. Use these capabilities to specify TC systems in terms of component counts, engineered structures, interfaces, problem spaces, etc.

## 2.3 Realizing Thermodynamic Computing

In this section we outline concepts to address the objective of realizing "Stage 2" thermodynamic computing systems (see TC Roadmap in Section 1.5) in which thermodynamic components representing evolving state variables are coupled to classical computing to form a hybrid computing system. Example thermodynamic components might include thermodynamic "bits," "neurons," "synapses," "gates," and "noise generators". We suppose that Stage 2 TCs will link these elements with traditional CMOS elements and an architecture focused on a particular class of problems.

**BACKGROUND AND SUMMARY OF THE CURRENT STATE**

There are already realizations of components employing underlying physics to extend beyond a typical purpose, such as storing a binary memory. As an example, oxide-memristors with complex non-equilibrium thermodynamics (Yang, Strukov, and Stewart 2013) have been designed to demonstrate multi-level switching, leading to the development of crossbar memristor arrays for fast and efficient vector-matrix multiplication. The training of these crossbars involves thermodynamically driven reconfiguration of the memristor components such as Ag migration in SiOx layers (Wang et al. 2017). Such devices have been incorporated into platforms for Hopfield networks in which the noise inherent in the memristors facilitates solutions to NP-hard problems by minimizing a mathematical problem function, similar to thermodynamic minimization (Kumar, Strachan, and Williams 2017). Another very different platform, namely the D-Wave series of platforms (https://www.dwavesys.com/quantum-computing), utilizes quantum processes (such as tunneling and adiabatic annealing) to perform energy minimization, though the algorithm can be realized using classical components as well. Yet another platform, the optical Ising machine (Fabre 2014), uses lasers and other optical components to perform minimization of a given problem's function. The aforementioned systems represent a handful of new directions in post-CMOS computing that overlap some of the themes and motivations of this report.

There are also several prior attempts to build prototypical computing systems based on thermodynamic principles. In 2005, Tanaka et al. (2005) developed an algorithm for designing multiple sequences of nucleic acids by considering minimum free energy. The resulting DNA computing algorithm reduced computation time drastically and is an early demonstration of thermodynamic computation using DNA biomolecules. Other examples of natural thermodynamic computing systems include the physarum polycephalum, a.k.a. slime mold, which has been "harnessed" to solve graph and other computational problems (Adamatzky 2010) and chemical reaction networks (CRNs). Another example is the recent demonstration of self-organization in a fully memristive neural network (Wang et al. 2018) enabled by diffusive memristors (Wang et al. 2017), which function based on electrochemical potential and interfacial energy minimization.

**LIMITATIONS OF THE CURRENT APPROACH**

In many ways we are at a crossroads similar to the one at the dawn of automated computation. In terms of the TC Roadmap (Section 1.5) we are approaching Stages 1A-B, but we do not yet have model systems to express TCs in general. Rather, we have a small set of interesting and suggestive devices and architectures but lack a unifying, general model. The paradigm for performing computation by TCs is still to be articulated, and the practice of designing such systems is still to be created. Among other needs, this foundational void calls collecting concepts and capabilities related to computation, applications, devices, and physics in pursuit of a more cohesive paradigm, engineering framework, and development plan.

In order to illustrate these limitations, consider that the components of classical computing systems with certain modifications might also serve as the thermodynamic components of Stage 2 systems, and that these modifications might be straightforward if a cohesive paradigm for TC existed. For example, CMOS transistors used to construct memories and logic gates also might be used to construct thermodynamic bits and neurons that leverage their inherent randomness in a "search" for a stable state or output. These classical and thermodynamic CMOS components might also be integrated with memristive components to create systems that fluctuate and stabilize organization over multiple spatial and temporal scales. In contrast, today's computing systems are designed to eliminate fluctuations, to separate scales, and to organize themselves according to an externally supplied program.

An additional limitation is that current mathematical representations of computing devices and systems fail to adequately describe thermodynamic properties. For example, component models focus on classical circuit quantities like current and voltage, while lacking descriptions of thermodynamic ideas like fluctuations and phase transitions. Although there have been recent realizations of the need for thermodynamics in the description of device and system behavior (Kumar and Williams 2018), the understanding is insufficient to support TC system development. Also, while it seems likely that new materials will be employed in the thermodynamic components like memristors, we lack the ability to model the effects of new materials on device function. In short, we need abstractions and models for thermodynamic effects across materials, devices, circuits, and systems.

**INSIGHT AND HYPOTHESES TO ADDRESS THESE LIMITATIONS**

While we lack a complete paradigm, we also foresee the ability to develop of Stage 2 TC systems with both practical and conceptual benefit around the current state of understanding. For example, because state of the art transistors are already in the mesoscale regime in which fluctuation effects are observable we can develop mathematical models of these effects and use them in the development of thermodynamic components for Stage 2 systems. We could also undertake the design of a thermodynamic memory by providing a fluctuating state to the classical components of a Stage 2 system that can be stabilized through interaction with it. Similarly, we might use combinations of thermodynamic transitive and memristive components to create neuron and synapse models to support machine learning models specified by the classical computing components. Certain well-known problems in computation that benefit from stochastic search (e.g. NP problems) might be targets for development of TC architectures. A TC architecture focused on deep neural network models, for example, could be highly homogeneous and non-hierarchical, as could be an architecture for simulating the Navier Stokes equation. On the other hand, architectures for systems to solve combinatorial optimization problems, such as restricted Boltzmann machine architectures, could be hierarchical but have their degree of hierarchy (branching ratios, etc.) tuned to the particular optimization problem.

Existing computing frameworks may also serve as jumping off points for the realization of Stage 2 TCs. For example, pyTorch and TensorFlow allow users to create artificial neural networks, and Brian, CARLsim, and PyNN allow users to specify biological neural networks. In all cases, a description language allows the user to specify essential elements (e.g., nodes, connections, and objective functions) and we suppose that such languages that might be expanded and generalized for TCs. These tools might then be used to describe a larger system, such as the behaving agent model illustrated in Figure 8, in which the TC evolves to predict outcomes, reduce surprise, and maximize efficiency as constrained by the model description.

The description language for the architecture given in Figure 8 would need to specify the input signals, the outputs of the system, and an optimization objective (i.e., a value system). The thermodynamic computer would then have to predict outcomes and maximize its fitness for the task at hand, with the least amount of energy necessary. While this example is not meant to be comprehensive, it illustrates some of the key ingredients needed in both the architectural and programming considerations of a thermodynamic computer.

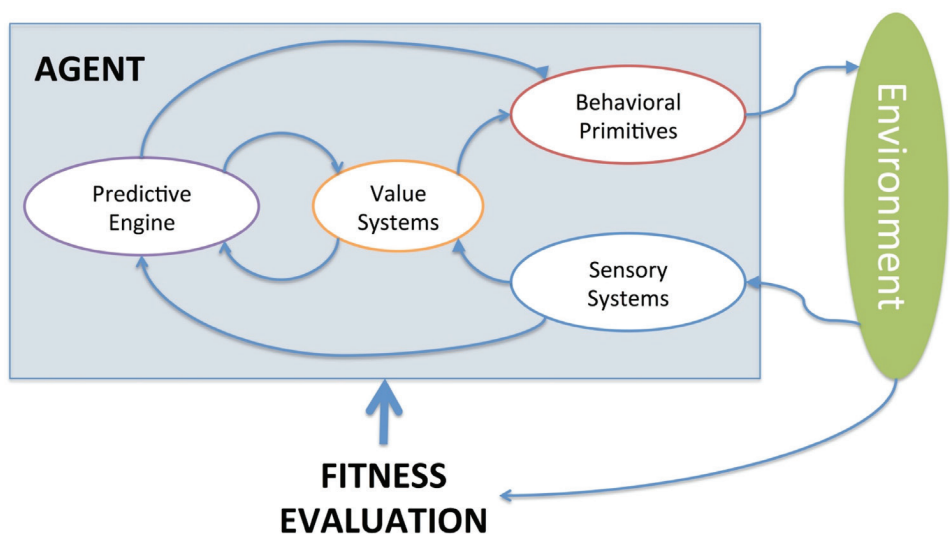

*Figure 8: Architecture for a thermodynamically efficient agent. The agent must take actions to maximize fitness in a complex, dynamic environment. It is endowed with innate values and behavioral primitives. It must evolve to predict outcomes that maximize positive value, minimize negative value, and reduce energy expenditure.*

The building of early stage components and systems are critical to understanding the challenges and potential of future large-scale systems. Efforts such as those just described would enable the evaluation of the various hypotheses developed in this report, such as the ability of TCs to self-organize (see Section 2.2, page 17), to leverage fluctuations, to operate at lower power, and to increase component density. The degree to which these hypotheses are correct would inform development and investment in TC in general. In the context of the current computing paradigm, the motivation for such early stage technology development is not just to pack more devices on a chip, but also to pack (potentially vastly) more functionality into that collection of devices through the development of a new computational paradigm.

**Scientific Challenges**

The realization of useful TCs will necessarily require the development of new understanding that converges and expands current understanding from the domains of thermodynamics, computing, and other disciplines. Various unsolved questions arise when considering how to realize functional TC architecture. We summarize these at the highest level as:

- Describing the new functionality thermodynamics can confer upon computing systems in addition to its well-known relevance to density and power efficiency *(what can a TC do?)*;
- Employing device and system thermodynamics to solve a problem *(how will we use a TC?)*;
- Characterizing the dynamics of complex TC architectures, even as its organization evolves *(how will a TC operate?)*;
- Developing abstractions and design principles for composable TC architectures comprising multiple thermodynamic elements *(how will we build a TC?)*; and
- Analyzing and quantifying the performance of a TC to guide further technological development *(how will we measure TC performance?)*.

**Research Directions**

While the scientific challenges of realizing TC systems can be stated in a relatively compact way, the research directions that might inform these challenges are diverse.

What follows is a list summarizing potential research directions at a high level:

- Develop mathematical models of materials for use in thermodynamic components.
- Develop and characterize novel materials for use in the fabrication of thermodynamic components.
- Fabricate and characterize thermodynamic components.
- Develop mathematical models of thermodynamic components.
- Develop methods for classical components to employ thermodynamic components as sources of stochasticity in probabilistic computations.
- Simulate TC architectures including both classical and thermodynamic components.
- Identify problem domains that are well fit to the current conceptual and technological capabilities of TC systems.
- Fabricate prototype TC architectures targeting identified problem domains and employing available components.
- Develop methods to characterize TC systems and extrapolate their effects in large systems: in particular, developing methods to characterize and extrapolate effects related to component density, power dissipation, fluctuation, stability, temperature and self-organization.
- Develop languages to specify and compose TC systems.

## 2.4. Applications of Thermodynamic Computers

In general, thermodynamic computation promises many orders of magnitude of speed up and efficiency over conventional computers for problems that involve optimization and intrinsically stochastic processes. Because their components and energy dissipation per operation could be many orders of magnitude smaller than conventional CMOS-based devices an exascale thermodynamic computer might fit in a laptop, as opposed to occupying a football field size building and requiring a dedicated multi megawatt power plant. Because they self-organize the complexity of the problem space addressed by the TC might be much larger and its solutions much more efficient. Also, there are already many examples of TC-like behavior in biology and bio-mimetic systems: for example, DNA computing, molecular error correction, kinetic proofreading, information-ratchet error correction, regulatory networks in cells, the brain, developmental biology, superorganisms (e.g. collective behaviors, flocking), and many more. But even if we can replicate behavior such as this in a thermodynamic computer, what problems will we use them to solve? In this section we speculate on future applications.

Artificial neural networks not only function in the presence of noise but noise can have a positive effect on the convergence of such networks to good solutions. Many other machine learning problems are tailor-made for thermodynamic computation. Boltzmann machines are an example of an explicitly thermodynamic model of computation: they learn by tuning the interactions between thermodynamic bits and optimize themselves by seeking the state of lowest free energy compatible with the statistics of the data that they wish to reproduce. Kernel methods such as support vector machines would benefit both from the larger size of the Hilbert spaces afforded by thermodynamic computation and from the intrinsically stochastic operation of the devices that allow a 'native' version of Monte Carlo sampling. Bayesian Neural Networks, genetic algorithms, and other metaheuristics similarly rely on access to efficient random samplers. If we extrapolate these ideas and consider building artificial neural networks the size of the brain (roughly $10^{12}$ neurons and $10^{15}$ synapses), an endeavor that would require orders of magnitude increase in energy efficiency, computational, and organizational capacity over current technologies, TCs may be the only feasible approach.

For example, consider the problem of simulating Navier-Stokes flows in microscopic systems where quantum effects and statistical fluctuations play an essential role. The stochastic dynamics of thermodynamic computers might be used to mimic these stochastic effects, so that thermal and quantum fluctuations in the computer itself can be injected into the computation to reproduce the thermal and quantum fluctuations in the simulated microfluidic flow. Similarly, at the level of simulations of macroscopic stochastic processes (like global atmospheric dynamics) TCs might have a natural advantage over traditional models of deterministic, noiseless computation. The number of cells in the discretized Navier-Stokes could be orders of magnitude larger than for conventional supercomputers (Lloyd 2000), and the stochastic nature of the computation need not be a hindrance for simulating systems that are themselves stochastic and chaotic.

Large-scale statistical mechanical systems are another source of inspiration and a potential class of applications for use on a TC due to their enormous range of complex behavior, including self-organization, Turing-complete dynamics, pattern formation, and highly correlated emergent phases. We expect that many of these aspects can be extended to TC's, offering a greatly increased range and sophistication of their computational capabilities, and, conversely, that TCs may be used to study these systems.

We note that many such systems are also being proposed as fruitful applications of quantum computing (Martonosi and Roetteler 2019)[8] because of the importance of coherent quantum effects such as entanglement at the microscale. TCs operating at subnanometer scale are themselves subject to strong quantum effects, which could be exploited to simulate small, partially coherent quantum systems. On the other hand, there is potential for hybrid machines with both quantum and thermodynamic components to simulate systems with large amounts of quantum coherence and entanglement by applying the better of the two models of computation. These advanced considerations are suggested as Stage 3 efforts in the TC Roadmap above.

---

[8] Next Steps in Quantum Computing: Computer Science's Role: https://cra.org/ccc/wp-content/uploads/sites/2/2018/11/Next-Steps-in-Quantum-Computing.pdf

# 3. Appendices

## Appendix A: Workshop Methods

The CCC Thermodynamic Computing workshop was held January 3-5, 2019 in Honolulu, Hawaii. It brought together a diverse group of physical theorists, electrical and computer engineers, computational biologists, and electronic/ionic device researchers with a strong understanding of thermodynamics in order to define a novel method of computing based around open system thermodynamics.

The workshop was divided into three primary components of thermodynamic computers: theory, physical systems, and model systems. Each of these areas was introduced through a tutorial presentation or panel – Gavin Crooks (Rigetti Quantum Computing) on theory, Joshua Yang (UMass Amherst) on physical systems, and the panel of Jeff Krichmar (UC Irvine), Suhas Kumar (Hewlett Packard Labs), and Todd Hylton (UC San Diego) on model systems. After these tutorials participants were split into three breakout groups, corresponding with one of the three topic areas, and assigned to identify priority research directions (PRDs) within that area (for instance non-Von Neumann architectures within the physical systems domain).

Each group categorized PRDs in terms of scientific challenges, summary of research directions, scientific impact of success, and technological/societal impact of success. Once each group had presented their findings to all the workshop participants, the organizing committee highlighted the cross-cutting PRDs that existed between each breakout group. These cross-cutting PRDs became the basis of the next set of break out groups, which again identified the scientific challenges, summary of research directions, scientific impact of success, and technological/societal impact of success for the PRD in question.

After the PRD groups presented their findings, group discussion was held amongst all the workshop participants to identify any remaining gaps and evaluate the validity of the items presented. Following those discussions, participants broke up into small groups to begin writing the text that formed the basis of this workshop report. The list of workshop attendees can be found in Appendix B.

THERMODYNAMIC COMPUTING## Appendix B: Workshop Attendees

| First Name | Last Name | Affiliation |
|---|---|---|
| Alexander | Alemi | Google LLC |
| Lee | Altenberg | University of Hawaiʻi at Mānoa |
| Tom | Conte | Georgia Institute of Technology |
| Sandra | Corbett | Computing Research Association |
| Gavin | Crooks | Rigetti Quantum Computing |
| James | Crutchfield | University of California, Davis |
| Erik | DeBenedictis | Sandia National Labs |
| Josh | Deutsch | University of California, Santa Cruz |
| Michael | DeWeese | University of California, Berkeley |
| Khari | Douglas | Computing Community Consortium |
| Massimiliano | Esposito | University of Luxembourg |
| Michael | Frank | Sandia National Laboratories |
| Robert | Fry | Johns Hopkins University Applied Physics Laboratory |
| Natesh | Ganesh | University of Massachusetts, Amherst |
| Peter | Harsha | Computing Research Association |
| Mark | Hill | University of Wisconsin-Madison |
| Todd | Hylton | University of California, San Diego |
| Sabrina | Jacob | Computing Research Association |
| Christopher | Kello | University of California, Merced |
| Jeff | Krichmar | University of California, Irvine |
| Suhas | Kumar | Hewlett Packard Labs |
| Shih-Chii | Liu | University of Zurich and ETH Zurich |
| Seth | Lloyd | Massachusetts Institute of Technology |
| Matteo | Marsili | Abdus Salam ICTP |
| Ilya | Nemenman | Emory University |
| Alex | Nugent | Knowm Inc |
| Norman | Packard | ProtoLife |
| Dana | Randall | Georgia Tech |
| Peter | Sadowski | University of Hawaiʻi at Mānoa |
| Narayana | Santhanam | University of Hawaiʻi at Mānoa |
| Robert | Shaw | Protolife |
| Adam | Stieg | University of California, Los Angeles - California NanoSystems Institute |
| Susanne | Still | University of Hawaiʻi at Mānoa |
| Elan | Stopnitzky | University of Hawaiʻi at Mānoa |
| John Paul | Strachan | Hewlett Packard Labs |
| Christof | Teuscher | Portland State University |
| Chris | Watkins | Royal Holloway, University of London |
| R. Stanley | Williams | Texas A&M University |
| David | Wolpert | Santa Fe Institute |
| Joshua | Yang | University of Massachusetts, Amherst |
| Yan | Yufik | Virtual Structures Research, Inc. |

## Appendix C: Bibliography

**THERMODYNAMIC COMPUTING**# NOTES

**THERMODYNAMIC COMPUTING**

**30**

# NOTES

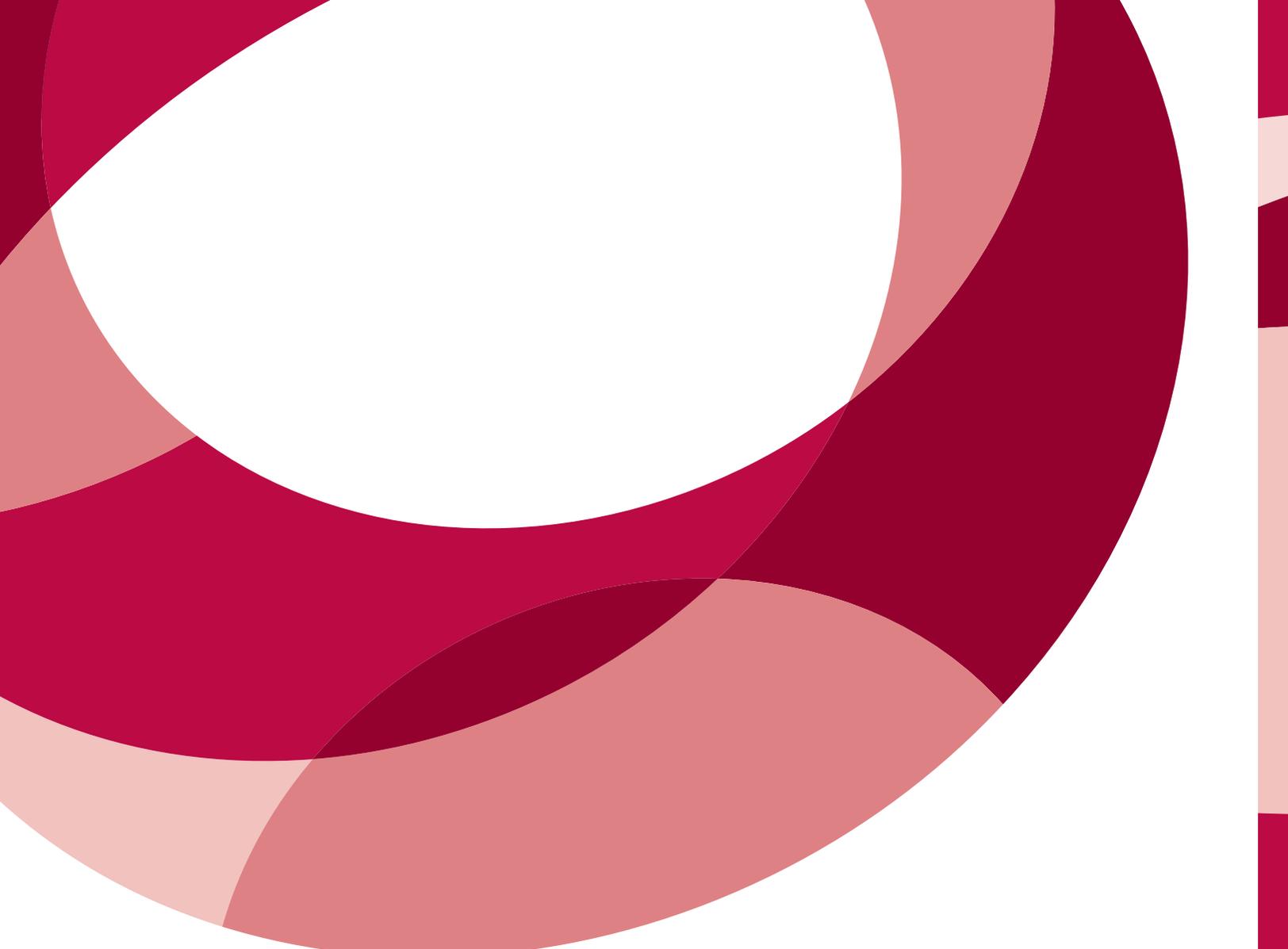

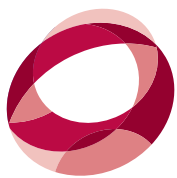

CCC
Computing Community Consortium
Catalyst

1828 L Street, NW, Suite 800
Washington, DC 20036
P: 202 234 2111 F: 202 667 1066
www.cra.org cccinfo@cra.org